# New Approaches and Trends in the Philosophy of Educational Technology for Learning and Teaching Environments[1]

Ismail Ipek [a], Rushan Ziatdinov [b,*]

[a] Department of Computer and Instructional Technologies, Istanbul Aydin University, Istanbul, Turkey
[b] Department of Industrial and Management Engineering, Keimyung University, Daegu, Republic of Korea
URL: http://www.ziatdinov-lab.com/

**Abstract**
The purpose of this study is to discuss instructional design and technology (IDT) model strategies for developing learning and teaching environments, based on philosophical approaches to educational technology theory. The study begins with a discussion of IDT models to define the history of educational technology or instructional technology theories, based on instructional strategies and improvements. In the study, authors discuss the strategies and steps that a design team should follow when designing learning environments in industry, business and military scenarios, based on the philosophy of educational technology and latest technologies, which should give way to effective learning environments. The steps include recognising terminology in educational technology concepts, psychological and instructional foundations in instructional design (ID), as well as approaches to educational technology. To recap, our purpose is to combine necessary IDT model strategies for the pedagogical design of learning environments, with new technologies. We will also discuss powerful IDT models that aim to meet the very high expectations of digital and humanist education. To develop a high-quality learning environment, we will explain technology design steps and practice in order to improve the learning of tasks, complex cognitive skills, attitudes, motivations and competencies in the future trends of educational technology. At the end of the study, integrated technologies in e-learning were discussed and presented, based on foundations of IDT and the philosophy of educational technology. These included pedagogical, technological and organisational technologies, as well as the main barriers of implementation, which, in turn, include the perspectives of students, teachers and designers, learning materials, digital education, epistemology in educational technology, courseware design, new technologies and contextual settings.

**Keywords:** philosophy of education, educational technology, instructional design, instructional design and technology, digital education

> *"I cannot teach anybody anything, I can only make them think."*
>
> *Socrates*

## 1. Introduction

Today, the developments in technology and the improvement of instructional design (ID) procedures with models have become focused on different sectors, such as industrial, business and military sectors, as well as on educational environments. With this in mind, new trends, technological innovations, philosophies pertaining to education, cultural perceptions and approaches all need to be discussed effectively, efficiently and globally, in order for us to recognise how badly we need international relationships, knowledge and active promotion of digital mobility with our partners for developing new schools, technologies, courseware, and learning design strategies, so that we may apply new ideas to them, and to potential learners. With this approach, scholars in the field of education, in addition to instructional design and technology (IDT), as well as other learning environments entailing different sectors, should commence discussing the philosophy of education from the beginning, in order to present possible changes for the future. They should start defining philosophical concepts, and follow, with understanding, the instructional terms and methods applied, thus far. Therefore, technology, education, instruction, learning design and multimedia design, for different levels, will be defined in order to provide the reaching of goals with learners and teachers, in learning environments for industrial, business and educational sectors.

The field of educational technology (ET) consists of both theory and ethical practice in the educational process, across different sectors (Januszewski & Molenda, 2008; Reiser & Dempsey, 2007; Seels & Glasgow, 1998; Seels & Richey, 1994). In this process, instructional design strategies provide contributions to global emerging technologies, for learning and teaching in educational technology and learning environments. These strategies deal with new technologies to develop learning environments, including digital learning, pedagogy online, learning design, humanism as digital humanism, collaborative learning, user-centered design and programming language, as well as instructional design models. The aim of the paper is to discuss and address basic dimensions, from past to present, for developing learning strategies to meet the objectives of temporary educational technology and its philosophical approach, which can be used interchangeably by instructional technology (IT) in the field of education, and in different sectors, as well.

## 2. Philosophy of Educational Technology and ID Models

All human societies, in the past and present, have focused on learning processes, and were interested in education. They have mentioned that teaching is the oldest career after dealing interests in education. There were no educated children who could read but, later on, they would learn to read, write and calculate, and also to employ the appropriate techniques in social learning, in order to reach individual goals. In addition to family life and learning skills, children need more activities and ways to live and thrive, within social groups. For this reason, perspectives on pluralistic societies might be changed, in individualistic terms, based on how the respective lifestyle changes. At this time, social life can be organised by rules, which are mentioned as parts of democracy and modern education. All social groups have been equally defined when meeting problems and shared responsibility, not based on majority, and have provided consensus with minority groups, as well. From this perspective, Dewey (1916) pointed out that the "primary ineluctable facts of the birth and death of the each one of the constituent members in a social group" make education important, despite the biologically inevitable fact that "the life of the group goes on" (p.3). The great importance of education is underscored, at this time. When a society is met by crisis, many problems emerge, which may be considered signs of educational breakdown, pertaining to components in educational technology, education, and educators, as well as technological changes with their effects on vocational education, industry, business, military, and their respective learning systems (Richey, Morrison, & Foxon, 2007).

As a result, philosophical trends and educational disciplines deal with ontology (existentialism), knowledge, epistemology and ethics, which, in turn, consist of ethics and aesthetics. All procedures and processes in educational movement for philosophers, from ancient times to the present day, include those of Socrates, Plato, Aristotle, Comenius, Locke, Rousseau, Pestalozzi, Voltaire, Diderot, Herbart, Dewey, Pazze, Skinner, Bacon, Bagler, Pablo Pierre and others (Phillips, 2008). Each philosopher or scientist of education addressed and worked on philosophical trends, which could be defined as idealism, realism, pragmatism, ontology, naturalism, behaviourism and analytical philosophy. In addition to these are today's educational technologies or instructional design dimensions, used to develop ID models for teaching in industry, business, military and schools, and may also be used in their e-learning environments with new technologies (İpek, İzciler & Baturay, 2008a, 2008b).

## 3. Psychological and Instructional Foundations of Instructional Design

Instructional design practices have been greatly influenced by a variety of theories concerning learning and instruction. Over many years, cognitive learning theory, behavioral learning theory, cognitive information processing theory, and Gagne's theory of instruction and instructional design, have had effects on instructional design and learning strategies. In recent years, schema theory, cognitive load theory, situated learning theory and constructivism have offered different approaches and learning strategies to develop methods for learning environments, as well as for design instruction, by using educational technology tools. Psychological foundations of instructional design are offered as philosophical perspectives for learning and instruction, in order to develop lessons by using those tools in educational technology. These approaches in learning design have defined and addressed the question of how to facilitate instruction, based on various theories, from behavioral theory to the constructivist approach.

Over the last decades of the twentieth century, constructivist epistemologies in learning sciences have been mentioned as alternatives to instructional sciences (Jonassen, Cermussa & Lonas, 2007). On the other hand, Hannafin & Hill (2007) discussed learning environments, including epistemological perspectives, design frameworks, and design practices. From this approach, epistemology was defined as the branch of philosophy concerned with the nature of knowledge, with the understanding of study, and ethical practices of facilitating learning and improving performance in educational technology. This point has already been presented by AECT[2] (Januszewski & Molenda, 2008). As discussed before, the word 'study' refers to research with all steps included, and ethics are not merely rules and expectations, but are a basis for practice and other stages, in the definition of ET. In addition to the main concepts of definition, facilitation considers the design of the environment, the resources and the provision of technological tools. For this, educational technology has a role for facilitating learning, rather than controlling. The epistemological perspectives are indicated as epistemological design frameworks and design practices. Thus, epistemological and psychological concepts have several interesting intersections, and invaluable contributions to the sectors of learning design, and the development of learning environments for the IDT field.

---

[2] Association for Educational Communication & Technology, Bloomington, IN, USA. URL: http://aect.site-ym.com/

The purpose of using educational technology is to enhance pedagogy, and to enable students to learn. A recent definition of educational technology is given by the Definition and Terminology Committee of AECT as follows:

**"Educational technology is the study and ethical practice of facilitating learning and improving performance by creating, using and managing appropriate technological processes and resources"** (Januszewski & Molenda, 2008).

With this definition, there are studies (research) and ethical practices for gathering information and analysing beyond traditional issues of research, and also for defining the field's ethical standards and case examples for practice. Educational technology also has a role in facilitating cognitive and constructivist learning theories, in order to obtain a connection between instruction and learning. It is a more facilitative approach, than a controlling one. Thus, educational technology claims to facilitate learning, rather than to cause or control learning activities. Facilitation includes design of learning environments, organising of resources, and the provision of learning tools. The learning environments can be in face-to-face settings, virtual environments, as well as microworlds, augmented realities, distance learning and learning designs.

Today, many instructional design models have been developed to be used within different learning environments for applying various learning theories in education, and for different purposes in different sectors. All ID models have generic steps of approach when it comes to the design of instructional systems, including design itself, development and evaluation stages. After all, later on, those stages have been referred to as analysis, design, development, implementation and evaluation stages. This was abbreviated as the (ADDIE) model. First, the generic IDI model, as a systematic design model, will be shown in the below image. The IDI model has three phases, which are shown in figure 1:

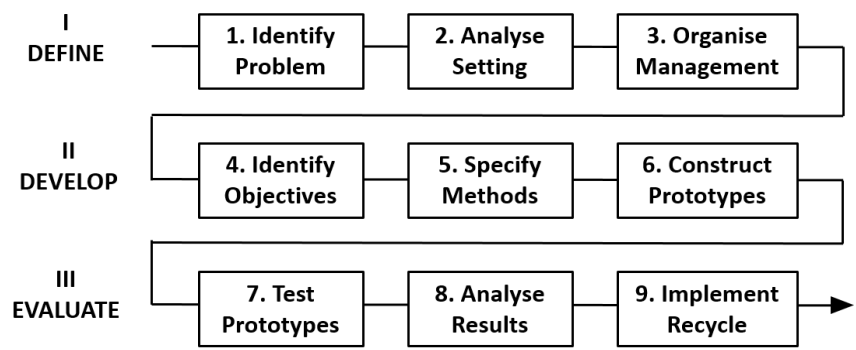

**Figure 1.** The Instructional Development Institute (IDI) Model

In addition to the IDI model, which is the first model-based approach, several ID models were developed after 1970. One of them is the Seels and Glasgow (1998) model, which is well known for using project development and class teaching, as well. The model is also available for cognitive and constructivist approaches to education for different sectors. The basic steps of the model, and a comparison to the generic ADDIE model, are given in table 1:

|   | ADDIE Model | Seels & Glasgow Model |
|---|---|---|
| 1. | Analysis | Need analysis, task and instructional analysis |
| 2.| Design      | Objectives and assessment, instructional strategy, delivery system selection and prototyping |
| 3.| Development | Materials development, formative evaluation                                            |
| 4.| Implementation | Implementation and maintenance                                                      |
| 5.| Evaluation  | Summative evaluation, diffusion and dissemination                                      |

**Table 1.** A comparison of the ADDIE model and Seels & Glasgow Model (1998)

Instructional design models include learning strategies, and learning performance does not use today what it connoted several years ago, when the first AECT definition, as learning procedure, was developed. There is a difference between retention of knowledge for testing purposes, and the acquisition of knowledge, skills, and attitudes used in the learning process. There are important variables to be found in learning, understanding, retention and assessment. For this reason, learning should be active, assessed by means of paper-and pencil tests, and should require demonstrations in keeping information, and in deep learning, as well.

**4. Future Instructional Design Trends and Newest Technologies**

Nowadays, instructional design trends and learning, as well as development professionals, should aware of what learners, clients and companies might be expecting. At this time, instructional designers and developers look at new ideas and reading materials for their own positions and those of their colleagues, and also look at industrial, business and military activities, for the development of future e-learning materials, in order for them to become much more productive and effective in different learning environments. Coffey (2017) indicated that instructional design trends, nowadays, include the following trends:

- Personalisation of instructional design, including traditional e-learning, social media, mobile phones, and the use of visual or technological landscapes.
- Providing contributions by means of training and communication, concerning the learning of corporate values and cultural changes.
- Consistent time with organisations and the cultural norms of learners, for rapid delivery of instruction.
- Using behavioral changes in ideal environments, and advances in blended and social learning, which can be designed and constituted from a combination of e-learning and integrated e-learning, with instructor-led content and social learning programs.
- Technological literacy and newer technologies will, in future, be part of instructional design and educational technology (IDT) for designers, developers and learners.

Finally, instructional design should be focused on how learners will actually take the training, and what most effective method of learning with new technologies will be. With this in mind, instructional design movements regarding educational technology or instructional technology, of course, enhances learning strategies in becoming successful for learning design with philosophical trends in educational technology and learning theory, as well as in instructional design theories and ID models. These trends influence problem-solving, motivation, achievement in sciences and social studies, visual and object-oriented programming behaviours for designing lessons, learning materials, developing e-learning environments for different sectors with instructional approaches and ID theories. In addition to these perspectives, there are other cases, such as digital education as pedagogy online, where teachers can be replaced with computer-based tutors, and where universities will be moving towards the hosting of online courses, in the next decade (Sharples, 2016). For this reason, we offer examples, such as massive open online courses (MOOCs) for future classrooms and learning environments. This process will be a form of modern education, supported by resources and services. There is also another point to consider, and that is that education can be designed for a new humanism and global university, by using information communication technology (ICT) applications. Thus, science and technology are becoming very important sections within the areas of humanism and educational technology, in which new, 21st century literacies are needed (Varis, 2017).

Based on emerging trends in instructional design and technology (IDT) approaches, the following print or digital technologies can be used as new technologies for future classrooms, and to develop other learning environments. Jacobs & Dempsey (2007) indicated emerging technologies for the future, and classified those technologies within learning environments. The environments consisted of object-oriented distributed learning, artificial intelligence applications, as well as cognitive science and neuroscience contributions, and they are explained as follows:

- Object-oriented distributed learning includes linked objects, electronic training jackets and metadata tag (courseware design-SCORM) areas.
- Artificial intelligence applications include instructional system approach variables, data mining technologies, expert system model stages, intelligent tutoring systems and cognitive models.
- Cognitive science and neuroscience contributions deal with brain structure and associated neural activity, including psychomotor behavior, recall of information, and decision making processes.

On the other hand, there are additional discussions concerning future learning and digital education. In the case of digital textbooks for digital learning environments, the amount of digital materials used is increased, but experts in this area are only predicting that digital concepts will be about 28% of the total instructional materials used in 2017 (Kring, 2015). Learning analytics is generally focused on subjects in STEM, on helping educators and on measuring students' concept levels, by way of a multitude of formats. If the approach is used effectively, learning analytics can help bring early hints to the surface, that could, in turn, indicate a students' performance in faculty and teaching subjects, as soon as possible. It also draws new patterns and analysis techniques for both simple and complex contents, in different fields. Basically, the learning analysis includes following stages as descriptions, diagnostics, predictions and prescriptions in IDT applications. With this process, digital education, in the next 10 years, could have its own virtual reality, which would deal with mobile computing and internet use, for learning environments (University of Europe Laureate Digital, 2017). Virtual and augmented reality will play an important role in gaming and gamification as well. These are new technologies for the future, based on cybernetics and nanotechnology, and are able to provide aid programs, such as electronic paper and wearable technology, in instructional design.

As we know, virtual reality has been used to help students to learn as effectively and efficiently as possible, in courses such as biology and other sciences. We should be aware of the boundaries between real and virtual learning environments, while designing courses and materials therein, as actual demonstrations or simulations based on ID theories and learning needs. As discussed, regarding virtual reality, augmented reality has similar functions but with more limited possibilities than virtual reality. It might become a dominant factor in the coming years, in the IDT sector. The applications can be seen on smartphone and tablet platforms, for different scientific programs. It is also considered to have an important role in connecting between the virtual and physical worlds. Although virtual reality is a complete step into virtual environments, augmented reality facilitates both visual and physical worlds. Once these systems are purchased for schools, it would be possible to conduct tests and simulations on them in real time, and to make use of live applications to learn, both by doing and by being immersed in real and active learning environments, at the same time. This achievement will be possible by using future technologies in educational settings, as well as digital humanism, and will also become more useful and important in real time for enabling students.

## 5. Learning Sciences and Research in IDT

In the last century, behaviourism and cognitive psychology had entailed learning and teaching theories, as well as instructional design theories. In recent years, constructivism, besides other learning approaches, had been mentioned to influence the instructional design field, whereas constructivism is not a single theory in education, and its roots depend on historical developments with philosophers and educators from the era BC to the modern day. Some of them, as philosophers and scientists, have been indicated in this paper above. Learning sciences review learning from different sets of assumptions and scientific perspectives, than instructional sciences do, as is the case with instructional design and technology (Jonassen, Cermussa & Lonas, 2007). Like traditional instructional sciences, it is a theory based and focused on cognitive sciences. As a constructivist approach, learning sciences rely on other factors, such as cognitive anthropology, situated learning, ecological psychology, distributed cognition and Dewey's philosophy, rather than other information-processing theories. In this approach, the learner is intentional, active and ready to take responsibility for constructing all personal mental models. This approach, epistemologically, deals with different learning types, enhanced learning environments, and collaboration among learners. The approach also has connections to design, and makes use of quantitative research methods, in order to establish general theories for designing instructional materials. As a result, design research activities, in IDT, follow many stages. These progress from initial design, to problem solving or needs assessment, to evaluation and reporting on the final product, as in project development for ID models.

## 6. Epistemology and Design Environments

Epistemology is a branch of philosophy. In history, many educators and scientists have focused on learning and teaching procedures, and the development of theories in instructional design and technology. At this time, knowledge of philosophy, truth information, designing materials and also design practice perspectives, have been indicated. There are systems approaches, grounded and/or cognitive design practices that will be used in design environments. All perspectives will be integrated into technologies to develop virtual and real-time simulations. These activities, in philosophical perspectives, include types of future instructional designs, such as learning environments, digital designs, gamification, applications of virtual reality, cloud-applications, integrated e-learning, mobile learning design, interface design, motivation, visual designs, multimedia learning methods and the use of instructional design models and networks in education, for the future of learning.

## 7. Reflection on Learning Theories and IDT

The definition of educational technology, or the field of instructional technology, had been discussed and redefined as a movement with instructional design process, during World War II (Seels, 1989). When we look at 1994 and the newest definitions of AECT in 2008, two main points are addressed as theory and ethics practice with sub-domains, which were given above. With the arguments concerning the definition of IDT, Januszewski and Molenda (2008) provide a full discussion of the conceptual components of this definition. In so doing, they provide a history of the field and its developments, based on learning theories and on instructional theory. On the other hand, Richey (2008) indicated that there are instructional designs and development steps, and also found that creation is broad and includes design, development and evaluation stages. But performance improvement for interpretation was limited in the new definition of ET. In another study, Tennyson (2010) indicated that learning theories and instructional design in history were mentioned by educational psychologists. One of them is Dewey (1975), and another is Thorndike (1913). Dewey was working on a special linking science between theory and educational practice, and Thorndike investigated some golden learning rules to apply in the teaching process, and developed a body of ID principles that include task analysis and teaching methods based on different types of findings from students and research applications. Both Dewey and Thorndike deal with learner behaviours and the law of effect and exercises, and Dewey focuses on discovery of learning strategies, as learning by doing. After this, the improvements provide integration of technology and instruction. As an example, Skinner (1954) developed the science of learning as programmed instruction. Thus, after this, Skinner became a founder of conventional computer-based instruction. Robert Gagne, Leslie Briggs, and Robert Glaser (1962) followed studies to develop ID models from the behavioral approach to cognitive theory (Gustafson & Branch, 2007; Morrison, Ross & Kemp, 2004; Reiser, 2007, Smith & Ragan, 2005). After all, the rest of the studies have conducted instructional theory research, which set the wheels in motion for the development of ID models, as well as technology education from needs assessment and writing objectives for the evaluation of learner performance. By the 1990s, instructional design moved toward an integrated instructional design approach. That means that more than one ID model stage was selected and used to solve learning problems, at the same time, by way of both learning theory and educational technology. At this time, the constructivist approach was accepted to design learning environments with reality and world experiences, and it then became a well-known paradigm. So, educational technology or instructional technology definitions are still under construction by being used interchangeably with instructional design, as well. As a result, we can make a decision about the instructional design and technology (IDT) field that's making great progress in solving learning and technology resource problems, by using IDT foundations and the philosophy of ET, and adapting the latest technologies.

## 8. Conclusion and Future Work Recommendations

The definition of instructional design and educational technology, with different approaches in the learning process, will still be an important topic with new technological developments. The latest technologies, and their considerations for instructional technology, will be the main subjects for developing learning materials as e-learning, or as integrated e-learning and virtual learning environments, with new trends for online MOOCs, SCORM or face-to-face learning classrooms. For future design and learning environments, instructional design concepts and contexts should be used effectively and efficiently, by experts in educational technology or instructional technology. The philosophy of educational technology will still continue to enhance learning environments, with design and programming, in the field of education. Conventional ID approaches, with new ID models, based on learning theories and approaches, will still be in use for developing instructional materials, procedures and performances in industry, business and military. As a result, epistemology, artificial intelligence, and philosophies in instructional design process will make vital sense for the development of future learning environments in schools, or outside of classrooms, with the latest technologies in the educational technology field. They also enrich the development of future classrooms values, concerning learning and teaching theories, for both the learners and the users. Thus, instructional design never stops when it comes to developing learning and teaching strategies, and it also creates new, effective and efficient digital and virtual learning materials in learning environments, for training in all sectors around the World.